\documentclass{aastex}
\usepackage{spr-astr-addons}
\usepackage{url}

\RequirePackage{color}

\def\be{\begin{equation}}
  \def\ee{\end{equation}}
\def\bea{\begin{eqnarray}}
\def\eea{\end{eqnarray}}

\begin{document}

\title{Effect of an external interaction mechanism in solving agegraphic dark energy problems}
\slugcomment{Not to appear in Nonlearned J., 45.}
\shorttitle{Short article title}

\shortauthors{A. Aghamohammadi et al.}

\author{A. Aghamohammadi$^{a}$\altaffilmark{1}}  \and
\author{Kh. Saaidi$^{b}$\altaffilmark{2}}  \and
\author{A. Mohammadi$^{c}$ \altaffilmark{3}} \and
\author{H. Sheikhahmadi$^{b d}$ \altaffilmark{4}} \and
\author{T. Golanbari$^{b}$\altaffilmark{5}}
\author{S. W. Rabiei$^{b}$\altaffilmark{6}}
\affil{$^{a}$Sanandaj Branch, Islamic Azad University, Sanandaj, Iran.} \and
\affil{$^{b}$Department of Physics, Faculty of Science, University of Kurdistan, Sanandaj, Iran.} \and
 \affil{$^{c}$Young Researcher Club, Larestan Branch, Islamic Azad University, Lar, Iran.} \and
\affil{$^{d}$Young Researcher Club, Sanandaj Branch, Islamic Azad University, Sanandaj, Iran.}


\altaffiltext{1}{a.aqamohamadi@gmail.com }
\altaffiltext{2}{ksaaidi@uok.ac.ir }
\altaffiltext{3}{abolhassanm@gmail.com}
\altaffiltext{4}{h.sh.ahmadi@uok.ac.ir or @gmail.com}
\altaffiltext{5}{t.golanbari@uok.ac.ir}
\altaffiltext{6}{w.rabiei@gmail.com}

\begin{abstract}
Agegraphic dark energy(ADE) and New-ADE models have been introduced  as two candidates for dark energy to explain the accelerated expansion phase of the Universe. In spite of a few suitable features of these models some studies have shown that there are several drawbacks in them. Therefore in this investigation a new version of ADE and New-ADE are studied which can improve such drawbacks which appear in the ordinary ADE and New-ADE scenario. In fact we consider an interacting model of  scalar field with  matter and after re-deriving some cosmological parameters of the model, we find out the best fit for the model. Actually by finding the best fitting for free parameters of the model, we show that our theoretical results are in a good agreement with observational data.
\end{abstract}



\section{Introduction}
During two past decades, numerously observational data, such as Supernovae type-Ia (SnIa) {\citep{1, 1a, 1b, 1c}}, Cosmic Microwave Background {\citep{2, 2a, 2b}} and so on, faced scientists with this shocking fact that the Universe is undergoing an accelerated expansion phase.  Some people look for the source of this acceleration in the geometrical part of the Hilbert-Einstein action and have studied the modified gravity {\citep{25, 25aa, 26aa, 27aa}}. On the other hand some researcher believe that   the Universe should be dominated by an ambiguous kind of fluid with negative pressure, called dark energy, which is able to provide such an expansion. The cosmological observational data express that the universe includes $73\%$ dark energy, $23\%$ dark matter and only $4\%$ baryons, note that the contribution of radiation could be ignored against to the other components of the Universe. The nature and origin of dark energy is unknown for scientists and this fact
 make this kind of fluid to one of the most puzzling aspects of the Universe. Up to now, many proposals have been introduced to realized this phenomenon. It seems that the best candidate for dark energy could be cosmological constant which has an equation of state as $\omega=p/\rho=-1$ {\citep{3, 3a, 3d, 3c,  3b, 3e}}. However the cosmological constant scenario suffers two well-known problems, namely the fine-tuning and cosmic coincidence problems {\citep{4}}. Some other proposals for dark energy are based on dynamical equation of state which are realized by scalar field mechanism. The idea has provided large classes of scalar field dark energy such as quintessence {\citep{5, 5a, 5c, 5b}}, k-essence {\citep{6, 6a, 6b}}, tachyon {\citep{7}}, phantom {\citep{8, 8a, 8b, 8aaa}}, quintom {\citep{9, 9a, 9b}}, and chameleon {\citep{10,11, 11a,17aa, 18aa, 19aa}}. \\
In the last few years, the models which include an interaction between  matter and other components of the model received more attention. In fact there are two different kind of interactions. One of them is an internal interactions between the components of matter, in which the field conservation relation is satisfied. In this kind, a term which describes this interaction is manually added to the model. The other one is an external interaction term between scalar field and matter. In this case the every conservation equation is modified. One type of the second kind is known as chameleon mechanism.
When a scalar field interact with matter (visible and invisible matter) through gravity or directly, produces a fifth force on the matter which may violate the weak equivalence principle (WEP) and creates a non-geodesic motion {\citep{Am, Be}}. There are some particular ways to trapping the fifth force effect. Some authors believe that the scalar field can be coupled differently to visible and invisible matter of the Universe. So according to this, for suppressing the effects of fifth force, they admit that the scalar field couples merely with the invisible matter {\citep{Caldera, Valiviita}}. Chameleon scenario is another mechanism to circumvent this force and solve the WEP problem, refer to {\citep{invis}} and references therein for more details.\\
Another proposal to probe the nature of dark energy is ADE model {\citep{12}}. Quantum mechanics in general relativity results in the uncertainty relation called Kar\`{o}lyhazy relation. In this scenario it is assumed that dark energy comes from both space-time and matter field fluctuations. Kar\`{o}lyhazy et. al. {\citep{13, 13a, 13b}} argued that the distance $t$ cannot be known better accuracy than $\delta t = \beta t_p^{\frac{2}{3}} t^{\frac{1}{3}}$ if we take $\beta$ as a dimension-less constant of order unity and $t_p$ as Planck time. Following this work, Maziashvili {\citep{14, 14a}} deduced that energy density of metric fluctuation of Minkowski space-time is given by $\rho_{d}\sim{1}/{t_p^2 t^2}\sim{m_p^2}/{t^2}$, where the quantity $t$ denotes the age of the Universe and is expressed as $t=\int_0^{t} dt$. One can realized that in ADE, the age of the universe is taken as the length measure (in contrast to the HDE which we take horizon as the length measure), which this selection in turn solve the casuality problem in HDE, however the origin ADE model suffers from a problem to describe the matter dominant epoch of the Universe. After that Wei and Cai introduced a new mechanism as New-ADE which the model uses a conformal time in Friedmann-Lema\^{\i}tre-Robertson-Walker (FLRW) space-time, known as $\eta=\int {dt}/{a}$, instead of time $t$. Therefore the energy density is given as $\rho={3m_p^2n^2}/{\eta^2}$, where $3n^2$ is a numerical constant {\citep{15}}. \\
Since New-ADE belongs to a dynamical cosmological constant, therefore one need a dynamical framework to consider it. One of such a framework is known as Brans-Dicke gravity. {Another dynamical framework which attains cosmologists attention is called chameleon model that first introduced by {\citep{10}} and {\citep{11,11a}}.} In this model, it is assumed that there is a scalar field which non-minimally is coupled to matter. The coupling causes the scalar field gets an effective mass which depends on the local matter density; indeed chameleon scenario is a way to get an effective mass for light scalar field via self interaction and interaction with matter. On the other hand, due to this coupling, the action and field equation is modified. Other important result is that the energy conservation equation is generalized as well. An interaction term appears in the right hand side of the relation which changes the behavior of matter density and equation of state parameter.\\
Both ADE and HDE have the same origin, however it is argued that ADE has a different IR-cutoff, and different IR-cutoff brings different results. In this work we would like to consider ADE and New-ADE in an interacting scenario. Therefore the effective dark energy in the model is taken equal to the energy density of ADE. The generalized conservation relation results in a equation of state parameter which has this ability to cross the phantom divide.  \\
The paper has been planned in the following form: In Sec.\;2 we derive the general form of the evolution equations and energy conservation equation; In Sec.\;3 the effective dark energy is taken equal to the ADE energy density and the behavior of dark energy density parameter, equation of state parameter, sound speed are obtained. In Sec.\;4 we take New-ADE as the effective dark energy and we study  the  behavior of cosmological parameter of the model. In Sec.\;5 we fit the data for ADE mechanism and find out the best fit for free parameters of the model and Sec.\;6 is devoted to the conclusion and discussion.
\section{Framework}
We consider the following action

\begin{equation}\label{1}
S = \int d^4x \sqrt{-g} \Big( \frac{1}{2}R - \frac{1}{2} \partial_{\mu}\phi\partial^{\mu}\phi - V(\phi) + f(\phi)L_m \Big),
\end{equation}
where $R$ is Ricci scalar, $G$ is Newtonian gravitational constant and   we take $8\pi G=1$.   $\phi$ is the scalar field with a potential $V(\phi)$ which has a non-minimal coupling with matter sector. The coupling is described in the last term of the action, where $L_m$ is the Lagrangian density of matter and $f(\phi)$ is an analytic function of  $\phi$.

To get the field equations one should takes variation of the action with respect to the independent variables. Taking variation of action with respect to the metric leads to the Einstein field equation as

\begin{equation}\label{3}
G_{\mu\nu}= f(\phi)T_{\mu\nu}+\Big[ \nabla_{\mu}\phi\nabla_{\nu}\phi - \frac{1}{2} g_{\mu\nu} (\nabla\phi)^2 \Big] - g_{\mu\nu}V(\phi),
\end{equation}
where $T_{\mu\nu}$ is matter energy-momentum tensor which is defined as
\begin{equation}\label{4}
T_{\mu\nu} = \frac{-2}{\sqrt{-g}} \frac{\delta\Big(\sqrt{-g}L_m\Big)}{\delta g^{\mu\nu}}.
\end{equation}
Here we merely assume that matter is a combination of (dark)matter and dark energy as  a perfect fluid which has a well-known energy-momentum tensor as $T_{\mu\nu}=(\rho+p)u_{\mu}u_{\nu}+pg_{\mu\nu}$, in which  $\rho=\rho_m+\rho_{\Lambda}$ and $p=p_m+p_{\Lambda}$.

Moreover to derive evolution equations a metric should be supposed to describe geometry of space-time, so we take an spatially flat case  ($k=0$) of  FLRW metric which is defined as

\begin{equation}\label{2}
ds^2=-dt^2 + a^2(t)\Big( dx^2+ dy^2+z^2 \Big),
\end{equation}
where $a(t)$ stands for scale factor of the Universe, $t$ is the cosmic time parameter.
By substituting the metric and the definition of matter energy-momentum tensor in the Eq.\;(\ref{3}), the Friedmann equation is given as follows

\begin{equation}\label{5}
3H^2=f(\phi)\rho + \frac{1}{2}\dot{\phi}^2 + V(\phi),
\end{equation}
\begin{equation}\label{6}
2\dot{H}+3H^2=-f(\phi)p - \frac{1}{2}\dot{\phi}^2 + V(\phi).
\end{equation}
On the other hand, taking variation of the action with respect to the scalar field $\phi$ leads to the following equation for scalar field

\begin{equation}\label{7}
\ddot{\phi}+3H\dot{\phi}=-V'(\phi)+f'(\phi)L_m.
\end{equation}
The prime denotes derivative with respect to scalar field and dot denotes derivative with respect to cosmic time $t$. One should specify the matter Lagrangian to more clarify the above equation. Based on {\citep{17, 17a, 17b, 17d, 17c}}, the Lagrangian of perfect fluid has two well-known definition as $L_m^1=p$ and $L_m^2=-\rho$, for the case when there is no any interaction between matter and other components of the model. But in this study where there is an interaction between matter and scalar field, the Lagrangian density degeneracy is broken. This means that the Lagrangian density $L_m^1=p$ and $L_m^2=-\rho$ has different results, so that according to {\citep{invis}} we pick out $L_m$ as $p$ which describe a geodesic motion for perfect fluid. \\
The next important equation is energy conservation equation. Using Eqs.\;(\ref{5}), (\ref{6}) and (\ref{7}) one could obtain the energy conservation equation for matter as

\begin{equation}\label{8}
\frac{d}{dt}{\Big(f(\phi)\rho\Big)}+3Hf(\phi)(\rho+p)=-p\dot{f}(\phi).
\end{equation}
As we expected the relation is not conserved that is due to the interaction of matter and scalar field. Since the energy density is a combination of dark energy and cold dark matter, namely $\rho=\rho_m+\rho_{\Lambda}$ and $p=p_m+p_{\Lambda}$, the conservation relation could be divided as following

\begin{equation}\label{9}
\frac{d}{dt}{\Big(f(\phi)\rho_{\Lambda}\Big)}+3Hf(\phi)(\rho_{\Lambda}+p_{\Lambda}) = -p_{\Lambda}\dot{f}(\phi),
\end{equation}
\begin{equation}\label{10}
\frac{d}{dt}{\Big(f(\phi)\rho_m\Big)}+3Hf(\phi)(\rho_m+p_m) = -p_m\dot{f}(\phi).
\end{equation}
The next energy density is related to the scalar field density that is indicated as $\rho_{\phi}$. According to the Friedmann equations, the energy density and pressure of scalar field could be defined as

\begin{eqnarray}
\rho_{\phi}&=& \frac{1}{2}\dot{\phi}^2 + V(\phi), \label{11} \\
p_{\phi}&=& \frac{1}{2}\dot{\phi}^2 - V(\phi), \label{12}
\end{eqnarray}
some manipulation lead one to a conservation relation for scalar field as

\begin{equation}\label{13}
\dot{\rho}_{\phi}+3H(\rho_{\phi}+p_{\phi})=\Big[ p_m + p_{\Lambda} \Big]\dot{f}(\phi).
\end{equation}
In this section we have obtained the required equation for our analysis. In the following sections we will consider the proposed model.
\section{ADE and scalar field}
 In comparison to the ordinary Friedmann equation, we define an effective dark energy as combination of dark energy $\rho_{\Lambda}$ and scalar field density as $\rho_{e\Lambda}=\rho_{\Lambda}+{\rho_{\phi}}/{f(\phi)}$, which $\rho_{e\Lambda}$ denotes effective dark energy. So, the Friedmann equation is rewritten as

\begin{equation}\label{14}
3H^2=f(\phi) \Big( \rho_m + \rho_{e\Lambda} \Big).
\end{equation}
An useful parameter in this study is energy density parameter $\Omega$. Here $\Omega_{e\Lambda}$ and $\Omega_m$ respectively will be taken equal to $\Omega_{e\Lambda}=f(\phi)\rho_{e\Lambda}/ \rho_c $ and $\Omega_m=f(\phi) \rho_m/ \rho_c$, in which $\rho_c$ is the critical energy density which is defined as $\rho_c=3H^2$. As a result, from the Friedmann equation we have $\Omega_{e\Lambda}+\Omega_m=1$.\\
To obtain energy conservation equations for effective dark energy,  using  Eqs.\;(\ref{9}) and (\ref{13}), one can achieve the following results
\begin{equation}\label{15}
\frac{d}{dt}{\Big(f(\phi)\rho_{e\Lambda}\Big)}+3Hf(\phi)(1+\omega_{e\Lambda})\rho_{e\Lambda} = \gamma \rho_m\dot{f}(\phi),
\end{equation}
\begin{equation}\label{16}
\frac{d}{dt}{\Big(f(\phi)\rho_m\Big)}+3Hf(\phi)(1+\gamma)\rho_m = -\gamma \rho_m\dot{f}(\phi),
\end{equation}
so that the effective pressure of dark energy is defined as $p_{e\Lambda}=p_{\Lambda}+{p_{\phi}}/{f(\phi)}$, and one has the effective dark energy equation of state parameter as $\omega_{e\Lambda}={p_{e\Lambda}}/{\rho_{e\Lambda}}$. Also $\gamma$ is the matter equation of state parameter which is defined as $\gamma={p_m}/{\rho_m}$. For $\gamma=constant$,   integrating of Eq.\;(\ref{16}) results in the following relation for cold dark matter energy density as
\begin{equation}\label{17}
\rho_m=\frac{\rho_{em}^0}{a^{3(1+\gamma)}f^{(1+\gamma)}(\phi)},
\end{equation}
where $\rho_{em}^0= f_{0}^{(1+\gamma)}(\phi)\rho_{m}^0$.
In this step, we suppose that the effective dark energy could be defined as ADE, in other word we assume that
\begin{equation}\label{18}
\rho_{e\Lambda}\equiv \rho_{ADE}=\frac{3n^2}{T^2},
\end{equation}
where $n$ is a numerical constant and $T$ is cosmic time and therefore $\Omega_{e\Lambda}$ is obtained as $\Omega_{e\Lambda}={f(\phi)n^2}/{H^2T^2}$.  Taking this assumption and using  Eq.\;(\ref{15}), the equation of state parameter of effective dark energy could be acquired as

\begin{equation}\label{19}
\omega_{e\Lambda}=-1 + \frac{2}{3}\frac{1}{n}\sqrt{\frac{\Omega_{e\Lambda}}{f(\phi)}} + \frac{\dot{f}(\phi)}{3Hf(\phi)} \Big( \gamma r - 1 \Big),
\end{equation}
where $r$ is ratio of cold dark matter and effective dark energy, namely $r={\rho_m}/{\rho_{e\Lambda}}={\Omega_m}/{\Omega_{e\Lambda}}$. The interaction term in this model generates an extra term for $\omega_{e\Lambda}$, which can justify the phantom divide line crossing.
By definition  an ansatz for $\omega_{e \Lambda}$, it can be considered as
\begin{equation}\label{19'}
\omega_{e \Lambda}+1=\omega_{0}+\omega_{1}(1+z)^{\beta}.
\end{equation}
\subsection{Data fitting}
In this section we want to fit the free parameters for ADE in an external scalar field interaction model. To achieve this purpose we use the $557$ Union $II$ sample database of $SnIa$, and  we assume $\rho_m= \rho_{radiation}+\rho_{baryon}+\rho_{dark matter} $. In this work the effect of  coefficient $f(\phi)$ which is appeared in  Eq.\;(\ref{5}), leads us to introduce some effective density parameters such as
$$\Omega_{e \Lambda}=f(\phi)\rho_{\Lambda}/\rho_{c},\qquad \Omega_{e m}=f(\phi)\rho_{m}/\rho_{c}.$$
Therefore in this case the Friemann equation is as
\begin{equation}\label{32}
3H^2=f(\phi) \Big( \rho_m + \rho_{e\Lambda} \Big).
\end{equation}
Combining Eqs.\;(\ref{15})-(\ref{18}), give
\begin{equation}\label{33}
3H^2=f(\phi) \Big(\frac{\rho^{0}_{em}}{a^{3(1+\gamma)}f^{(1+\gamma)}(\phi)} + \frac{3n^2}{T^2} \Big),
\end{equation}
where $\rho^{0}_{em}$ is the effective energy density of matter at the present time and $T$ is the cosmic time. Whereas the
$557$ Union $II$ sample database have collected from red shift parameter to various $SnIa$ , therefore we rewrite  $E=H/H_0$ versus $z$ as
\begin{equation}\label{34}
E^2=\frac{r_0(1+z)^3+(1+z)^{3\omega_0} \exp{\Big\{3\frac{\omega_1}{\beta}}[(1+z)^{\beta}-1]\Big\}}{r_0+1}.
\end{equation}
 To achieve the best-fit for free parameters we should compare the theoretical distance modulus, $\mu_{th}$, with observed, $\mu_{ob}$, of supernovae. In this model we have three free parameters $\omega_0$, $\omega_{1}$ and $\beta$. The distance modulus is defined as
\begin{eqnarray}\label{35}
\mu_{th}&=&5\log_{10}[D_{L}(z; \omega_0; \omega_{1}; \beta)]+\mu_{0},\\
\mu _{th}  &=& \tilde \mu _{th}  + \mu _0,\\
\end{eqnarray}
where $D_{L}(z; \omega_0; \omega_{1}; \beta)$ is given by
\begin{equation}\label{36}
D_{L}(z; \omega_0; \omega_{1}; \beta)=(1+z)\int_{0}^{z}\frac{1}{E(x; \omega_0; \omega_{1}; \beta)}d x.
\end{equation}
To comparing $\mu_{th}$ with $\mu_{ob}$ we need to obtain $\chi^{2}_{sn}$ which is defined by
\begin{equation}\label{37}
\chi^{2}_{sn}(\omega_0; \omega_{1}; \beta)=\sum_{i=1}^{557}\frac{[\mu_{th}(z_i)-\mu_{ob}(z_i)]^2}{\sigma_{i}^{2}}.
\end{equation}
\begin{eqnarray}\label{315}
 A& =& \sum\limits_{i = 1}^{557} {\frac{{(\tilde \mu _{th} (z_i ) - \mu _{ob} (z_i ))^2 }}{{\sigma _i^2 }}} , \\
 B& =& \sum\limits_{i = 1}^{557} {\frac{{\tilde \mu _{th} (z_i ) - \mu _{ob} (z_i )}}{{\sigma _i^2 }}} , \\
 C &= &\sum\limits_{i = 1}^{557} {\frac{1}{{\sigma _i^2 }}} , \\
 \chi^2 & =& A + 2\mu _0 B + \mu _0^2 C, \\
\end{eqnarray}
\\
In Fig.\;1, we show a comparison between theoretical
distance modulus and observed distance modulus of supernovae data. The red-solid
line indicates the theoretical value of distance modulus, $\mu_{th}$, for the best value of
free parameters $\omega_{1}=-1.65$, $\omega_{0}=1.1$ and $\beta=-2.25.$
\begin{figure}[h]\label{Fig5}
\centerline{\includegraphics[width=6cm]{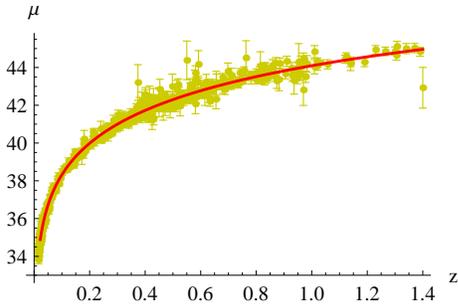}}
\caption{ {\small The observed distance modulus of supernovae (points) and the theoretical predicted distance
modulus (red-solid line) in the context of ADE model.}}
\end{figure}
A minimization of this expression leads to
\begin{eqnarray}\label{37'}
\chi^{2}_{sn_{min}}(\omega_0&=&1.1;~ \omega_{1}=-1.65; ~\beta=-2.25),\\
 \chi _{min}^2  &=& A - \frac{{B^2 }}{C}=542.75, \\
  \mu _0  &= & - \frac{B}{C}= 43.1089.
\end{eqnarray}
where implies ${\chi^{2}_{sn}}/{dof}$=
$\chi^{2}_{sn_{min}}/dof= 0.981 (dof = 553)$. This shows that this
model is clearly consistent with the data since $\chi^{2}/dof = 1$.
\\
Fig.2 show contour plots for the free parameters $\omega_{1}$ and $\beta$,  it is shown that the best value for these parameters are $-1.86<\omega_{1}<-1.62$ and  $-2.27<\beta<-0.73$ in which for  stability condition  $c^2 >0$, we have taken the interface between green and yellow sector, $\omega_1=-1.68$.\\
\begin{figure}[h]\label{Fig6}
\centerline{\includegraphics[width=6cm]{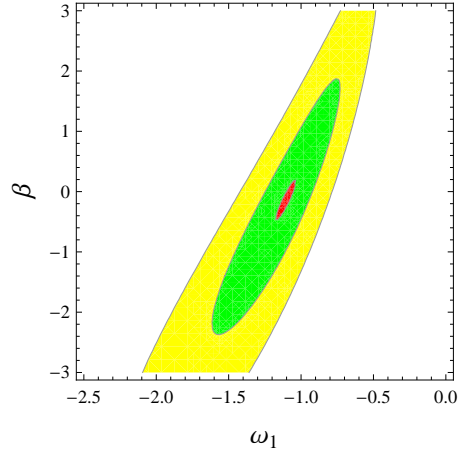}}
\caption{ {\small Contour plots for the free parameters $\omega_{1}$ and $\beta$, shows that the best value for these parameters are $-1.86<\omega_{1}<-1.62$ and  $-2.27<\beta<-0.73$.}}
\end{figure}\\
The evolution of effective dark energy parameter, $\omega_{e\Lambda}$, versus $z$, for $\omega_0= 1.1$, $\omega_1=-1.68$ and $\beta =-2.25 $ in the Fig.3 have been shown, It expresses that by growth $z$  the parameter get into the phantom phase.
\begin{figure}[h]\label{Fig1'}
\centerline{\includegraphics[width=7cm]{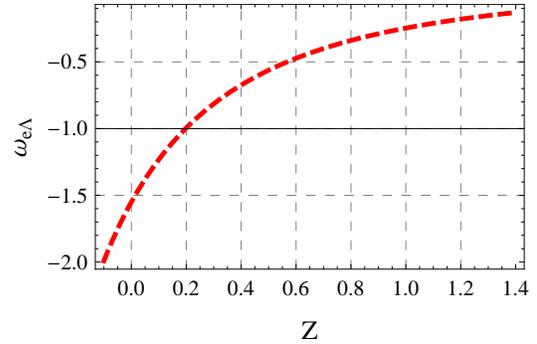}}
\caption{ {\small The plot shows the evolution of effective dark energy parameter, $\omega_{e\Lambda}$, versus $z$, for $\omega_0=1.1 $, $\omega_1=-1.68$ and $\beta =-2.25 $. }}
\end{figure}\\
where $\omega_{0}$, $\omega_{1}$ and $\beta$ are free parameters of the model which obtained from data fitting \citep{Campo}.
By using  Eqs.\;(\ref{15}), (\ref{16}) and (\ref{18}) one can obtain
\begin{equation}\label{19a}
f(\phi)=f_{0} t^{2} a^{-3\omega_{0}} \exp{\Big[3\omega_{1}\frac{{(z+1)}^{\beta+2}}{\beta+2}\Big]}.
\end{equation}
Here $f_{0}$ is the constant of integration.
Whereas $$\frac{\dot{f}(\phi)}{f(\phi)}=3H\big[\frac{2}{3tH} -\omega_{0}-\omega_1{(1+z)}^{\beta+2}\big].$$
Also evolution of effective dark energy parameter is investigated by  time derivative of $\Omega_{e\Lambda}={f(\phi)n^2}/{H^2T^2}$ as
\begin{equation}\label{21}
\Omega'_{e\Lambda}  =  -2\Omega_{e\Lambda} \Big[ \frac{\dot{H}}{H^2} + \frac{1}{HT} - \frac{\dot{f}(\phi)}{2f(\phi)H} \Big],
\end{equation}
where prime denotes derivative with respect to $N=\ln(a)$.\\
Using Friedmann Eq.\;(\ref{14}), energy conservation relations,    Eqs.\;(\ref{15}) and (\ref{16}),  and effective dark energy parameter, $\Omega'_{e\Lambda}$, is rewritten as
\begin{equation}\label{22}
\Omega'_{e\Lambda} =  3 \Omega_{e\Lambda}(1-\Omega_{e\Lambda})  \Big[ 1 -\omega_{0}- \omega_1{(1+z)}^{\beta} \Big].
\end{equation}
By integrating of this equation one can find out
\begin{equation}\label{23'}
\Omega_{e\Lambda}=\frac{D_{0}}{1+(1+z)^{3(1-\omega_{0})}\exp[\frac{-3\omega_1 {(1+z)}^{\beta}}{\beta}]}.
\end{equation}
Where $D_{0}$ is the constant of integration.
The behavior of effective dark energy density parameter has been plotted in Fig.\;4. It can be seen that by passing time the effective energy density parameter is saturated to $0.83$.
\begin{figure}[h]\label{Fig1}
\centerline{\includegraphics[width=7cm]{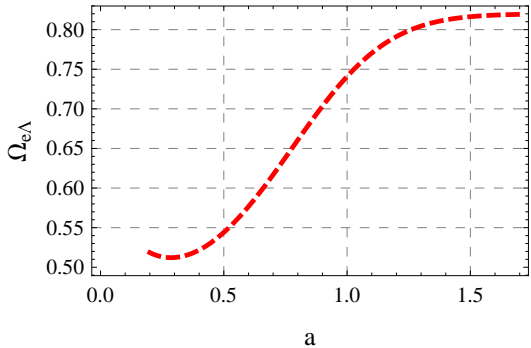}}
\caption{ {\small The plot shows the behavior of effective energy density parameter, $\Omega_{e\Lambda}$, versus $a$, for $\omega_0=1.1,\quad \omega_1=-1.68,\quad \beta=-2.25$.}}
\end{figure}\\
An significant  result of observational data is accelerated expansion of the Universe. A good cosmological model should be able to describe this acceleration. An useful quantity to investigate this property of the Universe is deceleration parameter which defined as $q=-1-{\dot{H}}/{H^2}$.
Using Eqs.\;(\ref{14}), (\ref{15}) and (\ref{16}), one achieves the deceleration parameter gives
\begin{eqnarray}\label{23}\nonumber
q&= -1+\frac{3}{2}\Big[1-\omega_0-\omega_{1}{(1+z)}^{\beta}\Big]\times\\ &
\Big(\frac{D_{0}}{1+(1+z)^{3(1-\omega_{0})}(t)\exp[\frac{-3\omega_1 {(1+z)}^{\beta}}{\beta}]}\Big).
\end{eqnarray}
It is clearly seen that for  $\omega_0=1.1,\quad \omega_1=-1.68,\quad \beta=-2.25$, (which have obtained from data fitting proseces)   $q<0$.

\subsection{Coincidence Problem}
In this subsection we want to consider one of the cosmological problems, namely coincidence problem. The problem addresses   that why the ratio of dark matter energy density, $\rho_m$, and dark energy density $\rho_{e\Lambda}$, is of order unity in present time. To indicate this ratio by $r$, we have
\begin{equation}\label{24}
r=\frac{\rho_m}{\rho_{e\Lambda}}=\frac{\Omega_{m}}{\Omega_{e\Lambda}}.
\end{equation}
To investigate the behavior of $r$, we use Eqs.\;(\ref{15}), (\ref{16}) and (\ref{19}) and therefore attain
 \begin{equation}\label{25a}
r=r_0(1+z)^{3(1-\omega_0)}\exp\Big[\frac{-3 \omega_1}{\beta}\big({(1+z)}^{\beta}-1\big)\Big].
\end{equation}
Where $r_{0}$ is the ratio at present time.
The behavior of coincidence parameter, $r$, has been plotted in Fig.5 which illustrates the variation of $r$ versus red-shift $z$. As one expect the plot shows that the parameter decreases by passing time. It means that as the universe is getting larger, the effective energy density overcome matter energy density. Based on the previous result we know that $\Omega_{e\Lambda}$ approach to $1$  means that $\Omega_m$ approach to zero, therefore the result of this section is consistent with the previous works.
\begin{figure}[h]\label{Fig2}
\centerline{\includegraphics[width=7cm]{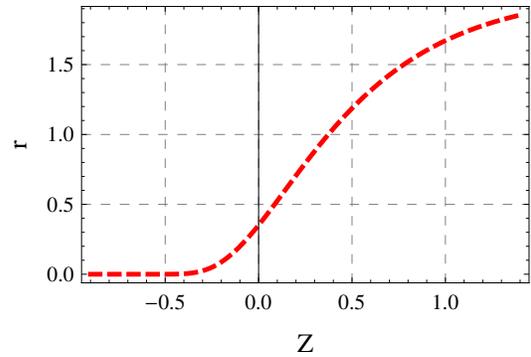}}
\caption{ {\small This figure shows the behavior of  $r$  versus $z$,  for $\omega_0=1.1,\quad \omega_1=-1.68,\quad \beta=-2.25$. }}
\end{figure}

\subsection{Square adiabatic sound speed}
Another interesting subject which could be addressed here is sound speed. This parameter is useful to investigate the classical stability of the models, therefore if sound speed be equal to a positive quantity the proposal can be considered as a viable model. In our model the squared adiabatic sound speed is defined as $c_s^2={dp_{e\Lambda}}/{d\rho_{e\Lambda}}$, where $\rho_{e\Lambda}$ and $p_{e\Lambda}$ are  the effective dark energy density and pressure respectively. Given; the equation of state and  effective dark energy parameters, $c_s^{2}$ could be expressed as follow
\begin{equation}\label{26}
c_s^2 = \omega_{e\Lambda}+\dot{\omega}_{e\Lambda}\frac{\rho_{e \Lambda}}{\dot{\rho}_{e \Lambda}},
\end{equation}
therefore using Eqs.\;(\ref{15})-(\ref{19a}), (\ref{26}) one can see
\begin{equation}\label{26'}
c_s^2 =\omega_{e\Lambda}+\frac{HT}{2}(1+Z)\frac{d\omega_{e\Lambda}}{dz}.
\end{equation}
Given, the equation $HT$ that is appeared in $c^2$ as following
 \begin{equation}
 HT=E(z)\Big[H_0T_o-\int_0^z\frac{d\acute{z}}{E(z)\big(1+z\big)} \Big],
 \end{equation}
  the behavior of $HT$  is plotted in Fig.6
\begin{figure}[h]\label{Fig4'}
\centerline{\includegraphics[width=7cm]{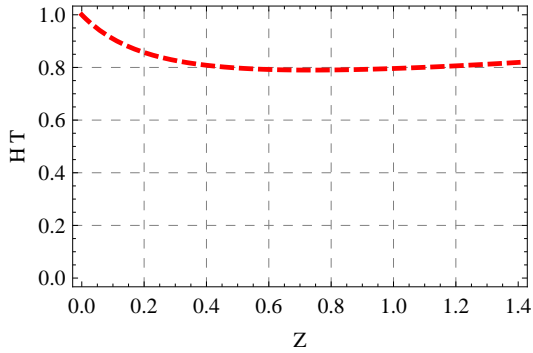}}
\caption{ {\small Behavior of $HT$ has been depicted versus red shift  $Z$.}}
\end{figure}\\
Hence, Fig.7, shows behavior of sound speed, $c^2$ versus red shift $z$. It is indicated that always $c^2>0$.
\begin{figure}[h]\label{Fig4}
\centerline{\includegraphics[width=7cm]{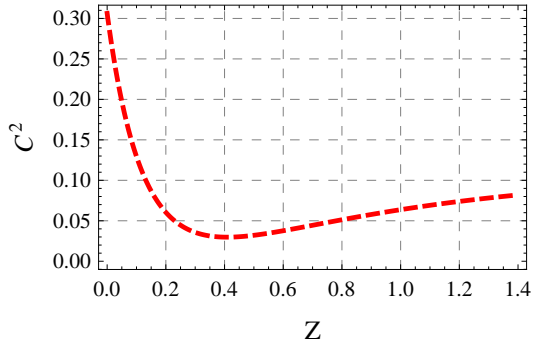}}
\caption{ {\small Behavior of sound speed $c^2$ has been depicted versus red shift  $Z$.}}
\end{figure}\\

As it is regarded by Fig.7,   this model solve the classical non stability problem of the original ADE mechanism.

\section{New-ADE and scalar field}
In New-ADE a conformal time, $\eta$, is substituted in the ADE relation instead of cosmological time $T$. So energy density is rewritten as $\rho_{NADE}={3n^2}/{\eta^2}$ where the conformal time is defined as $\eta=\int {dt}/{a}$. In this section we take the effective dark energy equivalent to the New-ADE, so we have
\begin{equation}\label{28}
\rho_{e\Lambda}\equiv \rho_{ADE}=\frac{3n^2}{\eta^2}.
\end{equation}
According to the previous section, the effective dark energy parameter is specified  as $\Omega^{\natural}_{e\Lambda}={n^2f(\phi)}/{H^2\eta^2}$. This definition of $\rho_{e\Lambda}$ and energy conservation relation,  Eq.\;(\ref{15}) lead us to derive the equation of state parameter as
\begin{equation}\label{29}
\omega^{\natural}_{e\Lambda}=-1 + \frac{2}{3a} \frac{1}{n}\sqrt{\frac{\Omega^{\natural}_{e\Lambda}}{f(\phi)}} -\frac{\dot{f}(\phi)}{3 H f(\phi)} .
\end{equation}
Using Eq.\;(\ref{19}) for this case leads to
\begin{equation}\label{29'}
\frac{\dot{f(t)}}{2Hf(t)}=\frac{1}{\eta H}-\frac{3\omega_0}{2}-\frac{3\omega_1{(1+z)}^{\beta}}{2}
\end{equation}\\
In comparison with the formal New-ADE and previous sections  there is an extra term and scale factor in the denominator of the second term, hence the model able to justify the easier phantom divide line crossing.
It is interesting to consider evolution of effective dark energy parameter as well. Doing the same process as the previous section lead one to the following equations
\begin{equation}\label{30}
{\Omega^{\natural}}'_{e\Lambda}=3 \Omega^{\natural}_{e\Lambda}(1-\Omega^{\natural}_{e\Lambda})  \Big[ 1+ (\omega_{1}-\omega_{0})+ \omega_1{(1+z)}^{\beta} \Big],
\end{equation}
therefore
\begin{equation}\label{31}
\Omega^{\natural}_{e\Lambda}=\frac{\Omega^{\natural}_{0}}{1+(1+z)^{3(\omega_0-1)\exp[\frac{-3\omega_1}{\beta}(1+z)^{\beta}]}},
\end{equation}
where $\Omega^{\natural}_0$ is the  constant of integration.
It easily can be realized that we have the same situation as the previous section. The differences come up due to appear of scale factor in the denominator of first term on the right hand side of the relation and also in a term in the relation of $\Omega^{\natural}_{e\Lambda}$. \\

On the other side, to specify the phase of the Universe, we consider deceleration parameter. Using Friedmann equation and energy conservation equations lead us to the following consequence for deceleration parameter
\begin{equation}\label{31}\nonumber
q= \frac{1}{2}+ \frac{3}{2}\omega^{\natural}_{e\Lambda}\Omega^{\natural}_{e\Lambda}
\end{equation}
The difference between Eqs.\;(\ref{31}) and (\ref{23}), is due to  $\Omega_{e\Lambda}$ and $\Omega^{\natural}_{e\Lambda}$. Therefore based on the our results the deceleration parameter gets a negative value and describe an accelerated expansion in agreement  with observational data. \\


\section{Conclusion}
In this work, an external interacting type of ADE and NADE has been picked out to consider the evolution of some cosmological parameters such as equation of state parameter, deceleration parameter, evolution of energy density parameter and sound speed. The general form of the evolution equations and energy conservation equation have been obtained. It is realized that in comparison to some other scalar field models, this model provides a generalized form of the energy conservation equation that expresses the right hand side of the relation is not equal zero. After that, the evolution equations and energy conservation equation have been rearranged to define an effective dark energy. Then the effective dark energy has been taken equal to the ADE and New-ADE to find out the behavior of mentioned parameters respectively. The acquired relations for equation of state parameter express that in contrast to the origin model, $\omega_{e\Lambda}$ could remain in phantom range for both cases of ADE and New-ADE, which in New-ADE this result happen easer due to the presence of the Universe scale factor in the denominator of some terms. Moreover we turned our attention to the deceleration parameter to obtain the expansion phase of the Universe in the model. The results represent that in agreement with observational data, the model could display an accelerated expansion phase for the Universe. \\
Also the evolution equation for dark energy density parameter was acquired in both cases of ADE and New-ADE. The results shown that by passing time, the density parameter increases and approaches to $1$ at late time, which expresses that dark energy overcomes matter density. From the relation it could be realized that when the parameter equal one, it ceases growing.\\
In addition, It was found out interesting to consider the behavior of $c_s^2$ (sound speed) and $r$ (the ratio of matter density and dark energy density) in the model. Therefore a shortly comments about these two subjects have been presented in this work.
In Fig.3 we have shown that the rate time of $r$ parameter smoothly decreases, and the square of sound speed could get positive value which have been explained in subsection.3.2 and is one of the advantages of the model. As well
 we have fitted the free parameters for ADE  in our model. To getting  this purpose we use the $557$ Union $II$ sample data-set of $SnIa$, to find the best fit for our models we combine radiation, baryonic and dark matter as matter component.
The best-fit for our model has shown that $\chi^2$ will be minimize for $\omega_{1}=-1.65$, $\omega_{0}=1.1$ and $\beta=-2.25$ and  this model has a good agreement with observational data as well.
\section{Acknowledgement}
The authors would like to thanks for anonymous referee for their helpful advices, which cause to improve the results of the model.


\begin{thebibliography}{}

\bibitem[Amendola.(2000,2004)]{Am} Amendola, L.: Phys. Rev. D {\bf 62}, 043511 (2000), Phys. Rev. D {\bf 62}, 103524 (2004)
\bibitem[Anisimov et al.(2005)]{9b} Anisimov, A.  Babichev, E.  Vikman, A.:  J. Cosmol. Astropart. Phys. {\bf 06}, 006 (2005)
\bibitem[Armendariz-Picon et al.(2000)]{6}Armendariz-Picon, C.  Mukhanov, V. F.  and Steinhardt, P. J.:  Phys. Rev. Lett. {\bf 85}, 4438 (2000)
\bibitem[Armendariz-Picon et al.(2001)]{6b} Armendariz-Picon, C.  Mukhanov, V. F.  and Steinhardt, P. J.: Phys.Rev. D {\bf 63}, 103510 (2001)
\bibitem[Astier(2006)]{1c} Astier. P, et al., Astron. Astrophys. {\bf 447}, 31 (2006)
\bibitem[Bean et al.(2001)]{Be} Bean, B. and Magueijo, J.: Phys. Lett. B {\bf 17}, 177 (2001)
\bibitem[Bennett et al.(2003)]{2a} Bennett. C. L, et al.: Astrophys. J. Suppl. {\bf 148}, 1 (2003)
\bibitem[Bicak et al.(1997)]{17}  Bicak, J. Kuchar, K. V.:  Phys. Rev. D {\bf 56}, 4878 (1997)
\bibitem[Brown et al.(1993)]{17a} Brown, J. D.  York, J. W.:  Phys. Rev. D {\bf 47}, 1420 (1993)
\bibitem[Brown(1993)]{17b} Brown, J. D.:  Class. Quant. Grav. {\bf 10}, 1579 (1993)
\bibitem[Cai(2007)]{12} Cai, R. G.:  Phys. Lett. B {\bf 657}, 228 (2007)
\bibitem[Caldera et al.(2009)]{Caldera} Caldera-Cabral, G. Maartens, R. and Schaefer, B. M.: JCAP {\bf 0907}, 027, (2009)
\bibitem[Caldwell(2002)]{8} Caldwell, R. R.:  Phys. Lett. B {\bf 545}, 23 (2002)
\bibitem[Caldwell et al.(2003)]{8a} Caldwell,R. R.  Kamionkowski, M. Weinberg, N. N.:  Phys. Rev. Lett. {\bf 91}, 071301 (2003)
\bibitem[Campo et al.(2011)]{Campo} del Campo, S. Fabriso, J .C. Herrerao, R. and  Zimdahl, Winfried.:  Phys. Rev. D \textbf{83}, 123006 (2011)
\bibitem[Carroll(2001)]{3c} Carroll, S. M.:  Living Rev. Rel. {\bf 4}, 1 (2001)
\bibitem[Chiba et al.(2000)]{6a} Chiba, T. Okabe, T. Yamaguchi, M.: Phys. Rev. D \textbf{62}, 023511 (2000)
\bibitem[Clemson et. al. (2009)]{5b} Clemson T. G, Liddle A. R, Mon. Not.Roy. Astron. Soc. {\bf 395}, 1585 (2009)
\bibitem[Cline et al.(2004)]{8b} Cline, J. M.  Jeon, S. Y. Moore, G. D.:  Phys. Rev. D {\bf 70}, 043543 (2004)
\bibitem[Einstein et al.(1917)]{3} Einstein,  A.:  Sitzungsber. K. preuss. Akda. Wiss. {\bf 142} (1917), [The prnciple
of relativity(Dover, New york, 1952), P. 177]
\bibitem[Elizalde et al.(2004)]{9} Elizalde, E.  Nojiri, S. Odinstov, S. D.:  Phys. Rev. D {\bf 70}, 043539 (2004)
\bibitem[Harko(2010)]{17c} Harko, T.:  Phys. Rev. D {\bf 81}, 044021 (2010)
\bibitem[Karolyhazy(1966)]{13} Karolyhazy, F.:  Nuovo. Cim. A {\bf 42}, 399 (1966)
\bibitem[Karolyhazy et al.(1982)]{13a} Karolyhazy, F.  Frenkle, A. lukacs, B.:  \emph{in Physsics as natural Phylosophy} edited by  Shimony, A. Feschbach, H.  MIT press, Cambridge, MA, (1982)
\bibitem[Karolyhazy et al.(1986)]{13b} Karolyhazy, F. Frenkle, A. lukacs, B.:  \emph{in Quantum Consepts in Space
and Time} edited by Penrose, R. Isham, C. J.  Clarendo Press, Oxford, (1986)
\bibitem[Khoury et al.(2004a)]{11} Khoury, J.  Weltman, A.:  Phys. Rev. Lett. {\bf 93}, 171104 (2004a)
\bibitem[Khoury et al.(2004b)]{11a} Khoury, J.  Weltman, A.:  Phys. Rev. D {\bf 69}, 044026 (2004b)
\bibitem[Maziashvili(2007a)]{14}  Maziashvili, M.: Int. J. Mod. Phys. D {\bf 16}, 1531 (2007a)
\bibitem[Maziashvili(2007b)]{14a}  Maziashvili, M.: Phys. Lett. B {\bf 652}, 165  (2007b)
\bibitem[Mota et al.(2004)]{10}  Mota, D. F. Barrow, J. D.:  Phys. Lett. B {\bf 581}, 141 (2004)
\bibitem[Nojiri et al.(2005)]{9a} Nojiri, S.  Odintsov, S.D. Tsujikawa, S.:  Phys. Rev. D {\bf 71}, 063004 (2005)
\bibitem[Nojiri et al.(2007)]{25aa} Nojiri, S.  Odintsov, S.D.: J. Phys. A. \textbf{40}, 6725 (2007)
\bibitem[Padmanabhan(2003)]{3e} Padmanabhan. T: Phys. Rep. {\bf 380}, 235 (2003)
\bibitem[Peebles et al.(1988)]{5}   Peebles, P. J. E.  Ratra, B.: Astrophys. J. Lett. {\bf 17}, 325 (1988)
\bibitem[Peebles et al.(2003)]{3b} Peebles, P. J. E.  Ratra, B.:  Rev. Mod. Phys. {\bf 75}, 559 (2003)
\bibitem[Peiris et al.(2003)]{2} Peiris. H. V, et al.: Astrophys. J. Suppl. {\bf 148}, 213, (2003)
\bibitem[Perlmutter et al.(1999)]{1a} Perlmutter. S. J, et al.: Astrophys. J. {\bf 517}, 565(1999)
\bibitem[Riess et al.(1988)]{5a} Ratra, B. Peebles, P. J. E.:  Phys. Rev. D {\bf 37}, 3406 (1988)
\bibitem[Riess et al.(1998)]{1} Riess. A. G, et al.: Astron. J. {\bf 116}, 1009 (1998)
\bibitem[Riess et al.(2004)]{1b} Riess. A. G, et al.: Astrophys. J. {\bf 607}, 665(2004)
\bibitem[Saaidi(2012b)]{invis} Saaidi, Kh.: arxive:1205.3542
\bibitem[Saaidi et al.(2011a)]{17aa} Saaidi, Kh.  Mohammadi, A.  and Sheikhahmadi, H.: Phys. Rev. D {\bf 83}, 104019 (2011a)
\bibitem[Saidi et al. (2012a)]{8aaa} Saaidi, Kh.   Aghamohammadi, A.    Sabet, B. and Farooq, O.:   Int. J. Mod. Phys. D. {\bf 21}, 1250057 (2012a)
\bibitem[Saaidi and Mohammadi(2012c)]{18aa}  Saaidi, Kh. and Mohammadi, A.: Phys. Rev. D  \textbf{85}, 023526 (2012c)
\bibitem[Saaidi et al.(2012d)]{19aa} Saaidi, Kh.  Mohammadi, A. Golanbari, T. Sheikhahmadi, H. and Ratra, B.: Phys. Rev. D 86, 045007 (2012d)
\bibitem[Saaidi and Aghamohammadi(2011b)]{27aa} Saaidi, Kh. Aghamohammadi, A.: Astrophys.Space Sci.
         {\bf 333}, 327 (2011b)
\bibitem[Saaidi et al.(2012e)]{26aa} Saaidi, Kh.  Vajdi, A. Rabiei, S. W. Aghamohammadi, A. and Sheikhahmadi, H.:
Astrophys. Space. Sci. (337), 739 (2012e)
\bibitem[Sahni et al.(2000)]{3d} Sahni, V. Starobinisky, A. A.:  Int. J .Mod. Phys. D {\bf 9}, 373 (2000)
\bibitem[Sen(2002)]{7} Sen, A.:  JHEP {\bf 0207}, 065 (2002)
\bibitem[Sotiriou et al.(2008)]{17d} Sotiriou, T. P.  Faraoni, V.: Class. Quant.: Grav. {\bf 25}, 205002 (2008)
\bibitem[Spergel et al.(2003)]{2b} Spergel. D. N. et al.: Astrophys.: J. Suppl. {\bf 148}, 175 (2003)
\bibitem[Steinhardt(1997)]{4} Steinhardt P. J, in critical problems in physics, edited by V. L. Fitch and D. R. Marlow
(Printed University Press, Prinston, NJ, 1997)
\bibitem[Valiviita et al.(2010)]{Valiviita} Valiviita, J. Maartens, R. and Majerotto, E.: Mon. Not. R. Astron. Soc. {\bf 402}, 2355, (2010)
\bibitem[Wand (1994)]{25} Wands, D.:   Class. Quant. Grav. \textbf{11}, 269 (1994)
\bibitem[Wei et al.(2008a)]{15} Wei, H. Cai, R. G.:  Phys. Lett. B {\bf 660}, 113 (2008a)
\bibitem[Wei et al.(2008b)]{18aa} Wei, H. Cai, R. G.:  Phys. Lett. B {\bf 663}, 1 (2008b)
\bibitem[Weinberg(1989)]{3a} Weinberg, S.:  Rev. Mod. Phys. {\bf 61}, 1 (1989)
\bibitem[Wetterich(1988)]{5c} Wetterich, C.:  Nucl. Phys. B {\bf 302}, 668 (1988)
\bibitem[Wu et al.(2007)]{18ab} Wu, X.  Zhang, Y.  Li, H. Cai, R. G.  Zhu, Z. H.  arXiv:0708.0349
\end{thebibliography}

\end{document}